\documentclass[pre,twocolumn,notitlepage,amsmath,amssymb]{revtex4-1}

\usepackage{graphicx}
\usepackage{float}
\usepackage{placeins}
\usepackage{mathtools}
\usepackage[bf]{subfigure}
\usepackage{xcolor}

\begin{document}

\title{Malleability of complex networks}

\author{Filipi N. Silva$^1$}
\email{filipinascimento@gmail.com}
\author{Cesar H. Comin$^2$}
\author{Luciano da F. Costa$^1$}
\affiliation{$^1$S\~ao Carlos Institute of Physics, University of S\~ao Paulo, PO Box 369, 13560-970, S\~ao Carlos, SP, Brazil\\}
\affiliation{$^2$Department of Computer Science, Federal University of S\~ao Carlos - S\~ao Carlos, SP, Brazil}


\begin{abstract}
Most complex networks are not static, but evolve along time.  Given a specific configuration of one such changing network, it becomes a particularly interesting issue to quantify the diversity of possible unfoldings of its topology.  In this work, we suggest the concept of \emph{malleability} of a network, which is defined as the exponential of the entropy of the probabilities of each possible unfolding with respect to a given configuration. We calculate the malleability with respect to specific measurements of the involved topologies.  More specifically, we identify the possible topologies derivable from a given configuration and calculate some topological measurement of them (e.g.~clustering coefficient, shortest path length, assortativity, etc.), leading to respective probabilities being associated to each possible measurement value.  Though this approach implies some level of degeneracy in the mapping from topology to measurement space, it still paves the way to inferring the malleability of specific network types with respect to given topological measurements.  We report that the malleability, in general, depends on each specific measurement, with the average shortest path length and degree assortativity typically leading to large malleability values. The maximum malleability was observed for the Wikipedia network and the minimum for the Watts-Strogatz model.  

\end{abstract}

\maketitle

\section{Introduction}

\begin{figure}[!htbp]
 \centering
\subfigure[]{\includegraphics[width=0.90 \linewidth]{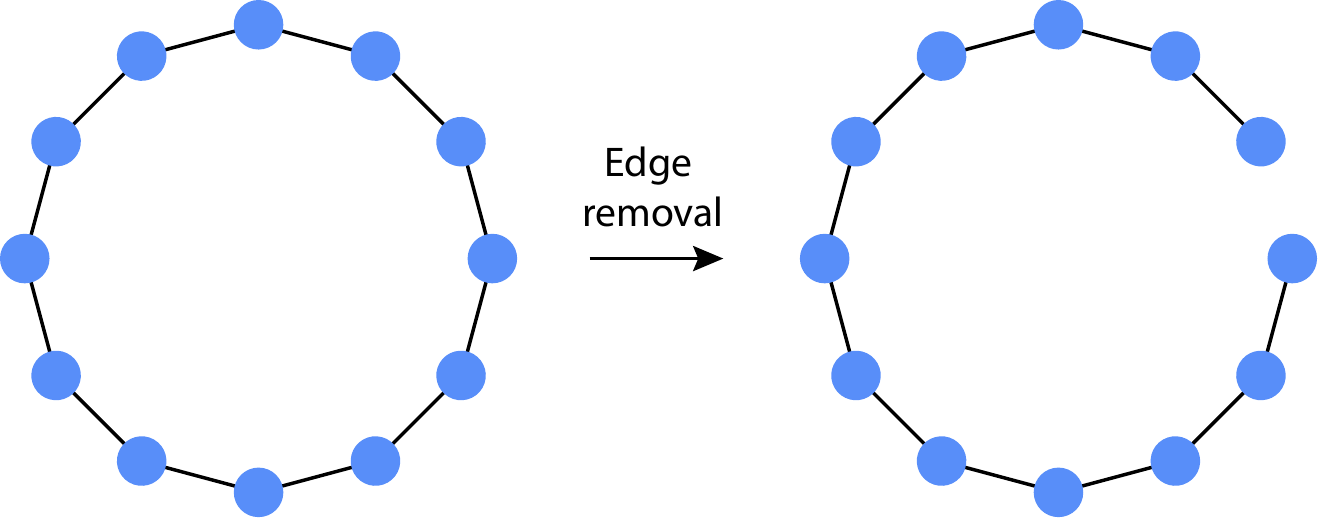}}\\
\subfigure[]{\includegraphics[width=0.90 \linewidth]{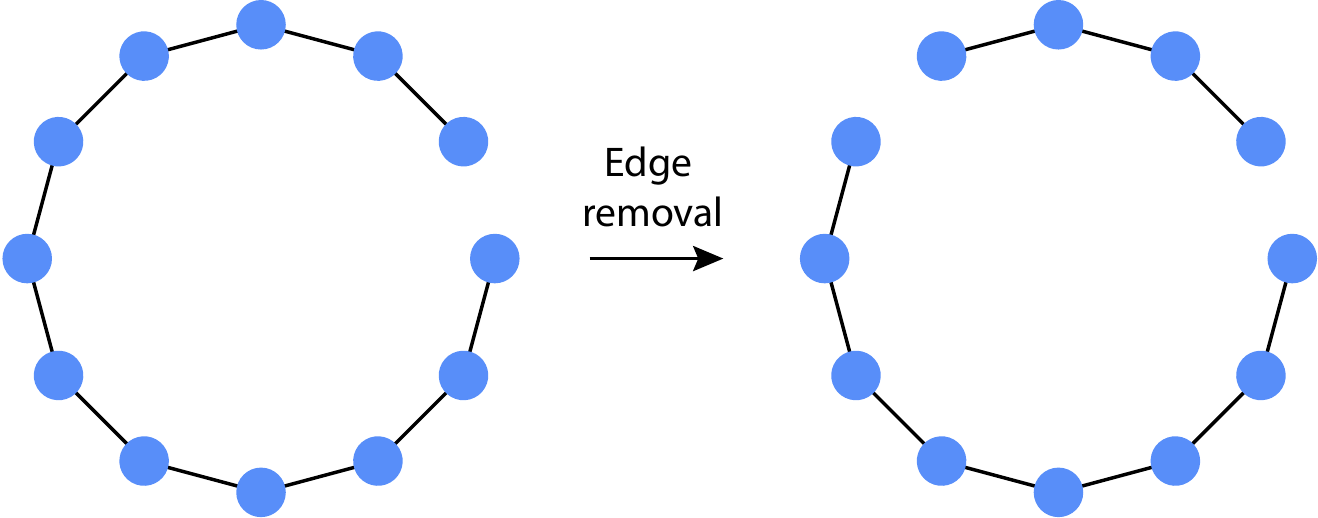}}
\caption{\label{f:ring} Example of a transformation from a ring network to a chain after the removal of any edge (a) and from a chain to two chains after subsequently removing any other edge (b).}
\end{figure}

Given a network subjected to possible changes on its topology, e.g.~ by adding or removing edges or vertices, an important question arises regarding the possible number of new network configurations induced by these topological changes.  For instance (please refer to Figure \ref{f:ring}), removing a link in a ring network will always imply in transforming it into a same chain (henceforth, all isomorphic configurations are considered as being identical).  However, removing a link from a chain network with $N$ nodes can yield $\lfloor N/2 \rfloor$ possible different networks.  These two simple examples make it evident that inducing small changes even to mostly similar networks can lead to rather different results, ranging from no change up to substantial structural modifications.  In this work, we refer to the effective number of possible new isomorphic topological configurations induced by the perturbation of a network as the \emph{malleability} of that structure.  

The malleability of a given network has several important theoretical and practical implications related to the characterization, robustness and adaptability of networks.  
In case a network is subjected to successive modifications, we can speak of its \emph{evolution} through successive \emph{topological states}, each having its own 
malleability.  So, an additional interesting question regards how the malleability of
a given network changes along subsequent topological modifications.  Will it be
kept constant, or increase/decrease along time?  What will be the sequence of topological
modifications capable of maximizing or minimizing the malleability of a given network?
Will the different histories of network structural modifications lead to rather
distinct scenarios regarding the respective malleabilities? 

For instance, let's consider the highway network of a given country or region at 
a given epoch along its topological evolution.  In case the respective network
malleability is large, it can be understood that the highway network can be
adapted to suit many different specific transit demands.  Otherwise, a small
malleability will imply that the aforementioned structure allows little chances of
being modified as desired, approaching a ``dead-end'' configuration.  Similar situations
arise in biology (e.g.~species co-evolution), computing (e.g.~a given distributed computer 
architecture), telecommunications (e.g. the Internet), and many other key real-world
systems and problems. 

These questions, to name but a few, are located at the core of network science studies 
for their theoretical and practical implications. For these reasons, it becomes important 
to derive a suitable definition of network structural malleability, and to perform 
studies aimed at characterizing several types of theoretical and real-world networks along 
their evolution.

The principal constraint to defining and studying network malleability concerns the
fact that several incremental topological modifications will produce isomorphic
network configurations, and these need to be identified and treated as being identical.
The problem is that isomorphism detection is a particularly demanding computational
task. In order to circumvent this limitation, we adopt an alternative approach in which
the networks are mapped into a measurement (or feature) space, so that each 
original network is described in terms of a set of its topological properties.
The rational here is that two isomorphic networks will necessarily yield identical
respective feature mappings.  Though this transformation from the topological space
into a given feature space is non invertible (i.e.~two networks having identical
features are not necessarily isomorphic), this approach will still be capable
of providing insights about the malleability of networks while avoiding the combinatorial
complexity implied by isomorphism identification.  This characterization will be all
the more accurate provided the selected set of features is capable 
of characterizing the topology in a more complete manner.  Interestingly, there is
an additional bonus of adopting this feature-based approach to network malleability
in the sense that we can now speak of the malleability of a given network \emph{with respect}
to specific topological properties.   For instance, we may find that a given network
or network model is more malleable regarding its assortativity than average shortest 
path.  Such results have good potential for better understanding the intrinsic
properties of model and real-world networks.

We start by presenting the definition of malleability in Section~\ref{s:malleability}. In 
Section~\ref{s:malleability_mea} we present the calculation of malleability with respect to
specific measurements. The networks and experiments applied for calculating the malleability
are described in Section~\ref{s:mal_exp}. In Section~\ref{s:results} we present the results, which
are discussed in Section~\ref{s:conclusions}.

\section{Malleability of complex networks}
\label{s:malleability}

\begin{figure*}[]
  \begin{center}
  \includegraphics[width=0.85\linewidth]{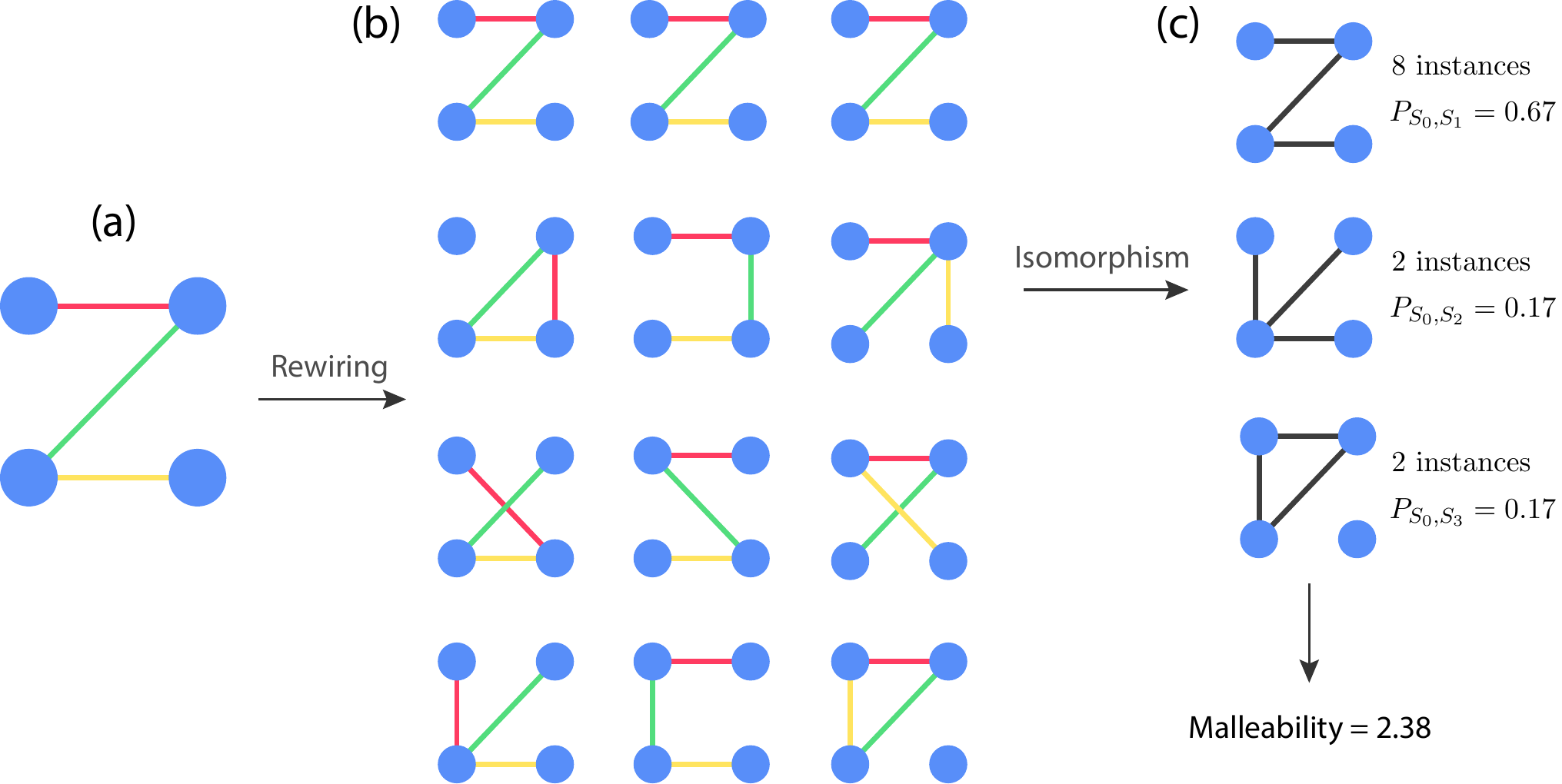} \\
  \caption{By rewiring an edge of the original network (a), $12$ new networks can be created (b). Notice that three of the networks are identical to the original. Among these $12$ networks there are only $3$ canonical forms, which are shown in (c) together with the number of times they appear in (b), as well as the corresponding probabilities. Such probabilities are used for calculating the malleability.}
  ~\label{f:motifs}
  \end{center}
\end{figure*}

\begin{figure}[]
  \begin{center}
  \includegraphics[width=\linewidth]{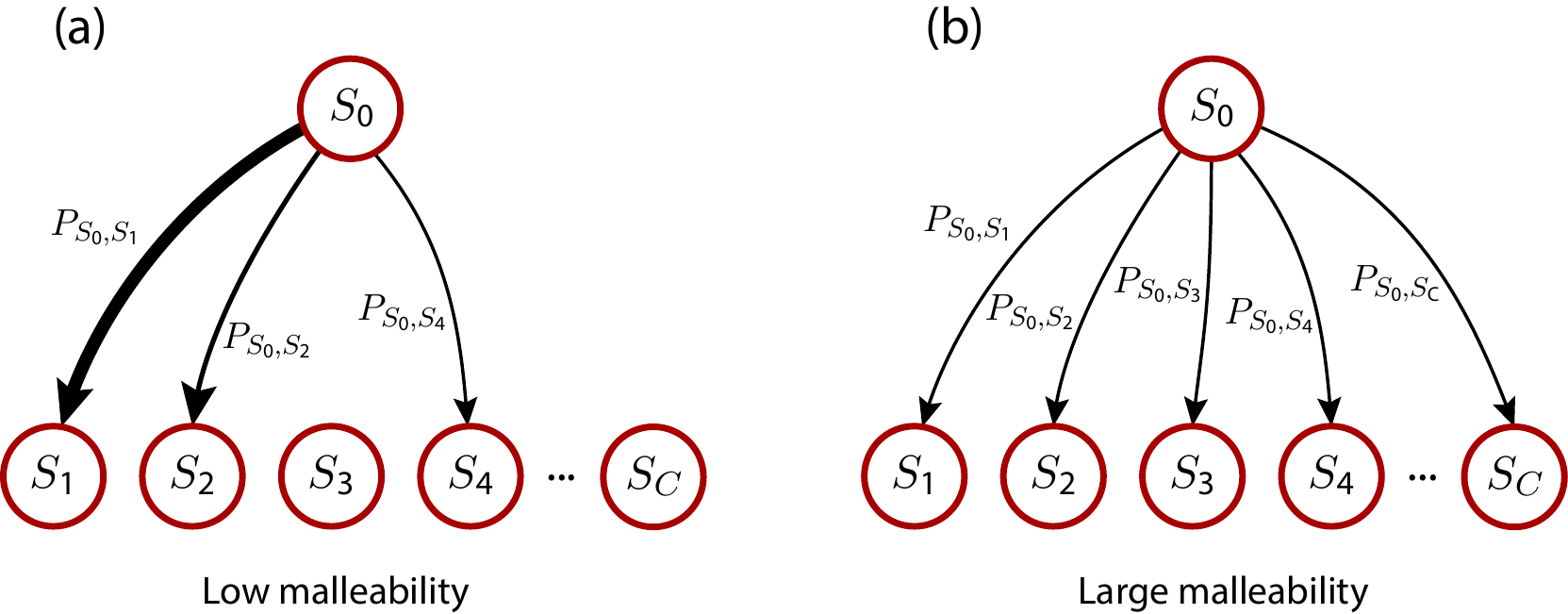} \\
  \caption{Example of states $S_0$ with distinct malleability. Thicker edges indicate larger transition probabilities. In (a) $S_0$ usually transitions to $S_1$, hence it has low malleability. In (b) $S_0$ can transition to all states with similar probability, which means that the state is more malleable.}
  ~\label{f:malleability}
  \end{center}
\end{figure}

Given a specific network (as illustrated in Figure~\ref{f:motifs}(a)) and a set of modification rules (e.g. a rewiring), a respective set of derivable network configurations is obtained (Figure~\ref{f:motifs}(b)). Many of these networks are isomorphic, yielding a smaller set of configurations $\mathcal{S}=\{S_1, S_2,\dots, S_C\}$, which are henceforth called states (Figure~\ref{f:motifs}(c)). Each state $S_i$ is reached with probability $P_{S_0,S_i}$, where $S_0$ corresponds to the original state of the network. If the states $S_i$ are reached with unequal probabilities (as shown in Figure~\ref{f:malleability}(a)), then we consider that the network has low \emph{malleability}. That is, at this state it has little potential for being modified. If, on the other hand, many distinct states are reached with equal probabilities (as shown in Figure~\ref{f:malleability}(b)), the network can be understood as being more malleable. The malleability also should increase with the number of reachable states $C$.

In order to quantify the malleability of the network, we first represent the transition probability from state $S_0$ to $S_i$ as $P_{S_0,S_i}$. Then, the entropy of the transition probabilities for the reference state $S_0$ can be defined as

\begin{equation}
\mathcal{E}_{S_0}=-\sum\limits_{S_i\in\mathcal{S}}P_{S_0,S_i}\log(P_{S_0,S_i}).\label{eq:entropy}
\end{equation}
The entropy indicates the heterogeneity of the distribution of transition probabilities.  By taking the exponential of the entropy, the following derived measurement becomes congruent with
the effective number of states

\begin{equation}
A_{S_0}=e^{\mathcal{E}_{S_0}}.\label{eq:accessibility}
\end{equation}
This quantity is equal to one when only a single state can be reached from $S_0$. When all reachable states are accessed with equal probability, the malleability becomes equal to $C$. 


\section{A methodology for malleability estimation}
\label{s:malleability_mea}

Ideally, it would be interesting to calculate the malleability in the original transition space, as illustrated in the previous section.  However, such an approach may require a high computational cost because of the combinatorial complexity implied by the many ways in which a graph can be modified.  In the following, we present a methodology for approximating the transition probabilities by defining a measurement space.  This approach, however, implies a shortcoming in the sense that the mapping from the states in the original space to those in the feature space will not necessarily be one-to-one.  Such a degeneracy is a consequence of the intrinsic lack of discriminability of most measurements (e.g. many nodes in a network can have the same degree).  As a consequence, a state in the feature space can correspond to two or more states in the original, complete transition space of each network.  On the other hand, the degree of degeneracy can be reduced by taking several complementary measurements into account, which has the bonus of allowing the immediate characterization of several properties of the analysed networks.

We begin by defining a \emph{measurement space} $\mathcal{M}=M_1\times M_2\times\dots\times M_l$, where $M_k$ is a topological property, such as the average degree or assortativity. A network state can then be represented as a point $\tilde{S}_i=(m_1, m_2, \dots, m_l)$ in this space. The transition probabilities to the reachable states can be calculated, and we can apply Equations~\ref{eq:entropy} and~\ref{eq:accessibility} to find the malleability of the network. Here we aim at understanding the malleability in terms of specific measurements. Therefore, a malleability value is derived for each measurement in $\mathcal{M}$. It is important to note that the malleability of distinct measurements can vary greatly, for instance, if we choose the average degree and a edge deletion dynamic, the resulting malleability will always be $1$ while for other measurements, such as average clustering coefficient, it can attain higher values.

Estimating the transition probabilities can be done in a number of different ways. Here we define an algorithm aimed at iteratively estimating such probabilities with respect to a specific measurement.

Given a network state $S_0$, a measurement $M$, a modification rule $R$, the number of trials $N$ and an empty list $L$, the following procedure can be applied to estimate the malleability
\begin{itemize}
\item {\bf (1)} apply the modification rule $R$ on $S_0$, generating a state $S_i$
\item {\bf (2)} calculate the adopted measurement $M$ for $S_i$ and store the result into $L$
\item {\bf (3)} repeat steps {\bf (1)} and {\bf (2)} $N$ times
\item {\bf (4)} identify the unique values $\tilde{S}_i$ in $L$
\item {\bf (5)} calculate their relative frequencies, corresponding to the respectively estimated transition probabilities $\tilde{P}_{S_0,\tilde{S}_i}$
\item {\bf (6)} apply equations~\ref{eq:entropy} and \ref{eq:accessibility} to $\tilde{P}_{S_0,\tilde{S}_i}$
\end{itemize}
Note that the unique values found in step (4) are calculated up to the machine precision.

\begin{figure*}[]
  \begin{center}
  \includegraphics[width=\linewidth]{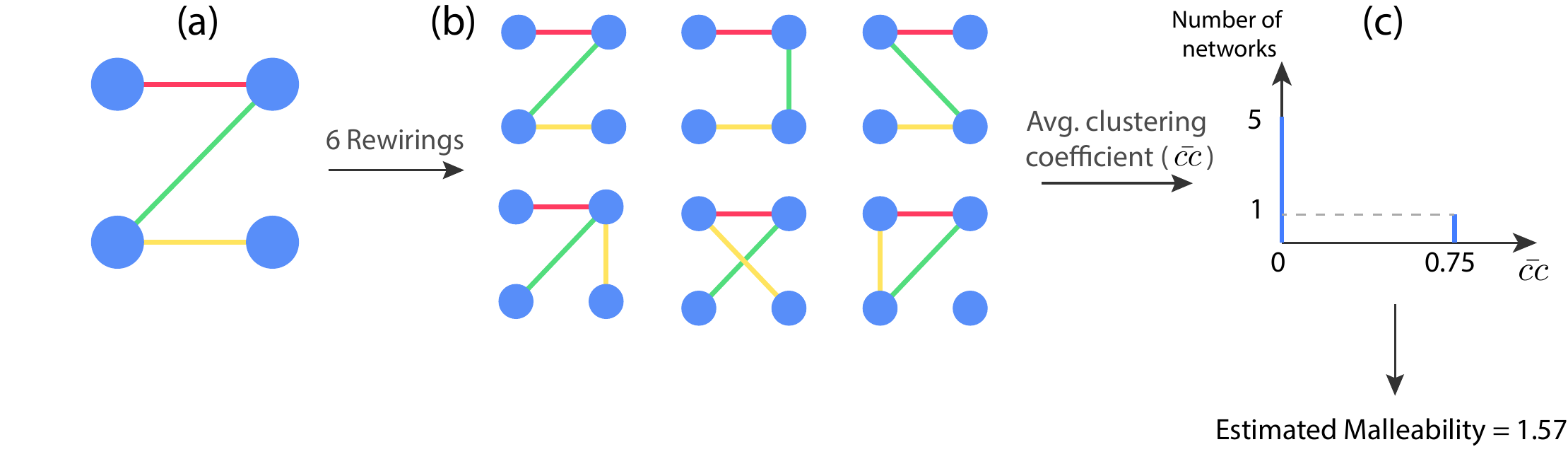} \\
  \caption{Illustration of malleability estimation. The original network (a) is rewired 6 times, generating $6$ networks (b). The average clustering coefficient of each network is calculated, and the unique values, up to the machine precision, and their respective frequencies are obtained (c). The malleability for the clustering coefficient of this particular network can then be estimated.}
  ~\label{f:motifsApprox}
  \end{center}
\end{figure*}

Figure~\ref{f:motifsApprox} show an example of malleability estimation for the same network considered in Figure~\ref{f:motifs}.

\section{Experiments}
\label{s:mal_exp}

\begin{figure*}[]
  \begin{center}
  \includegraphics[width=0.80\linewidth]{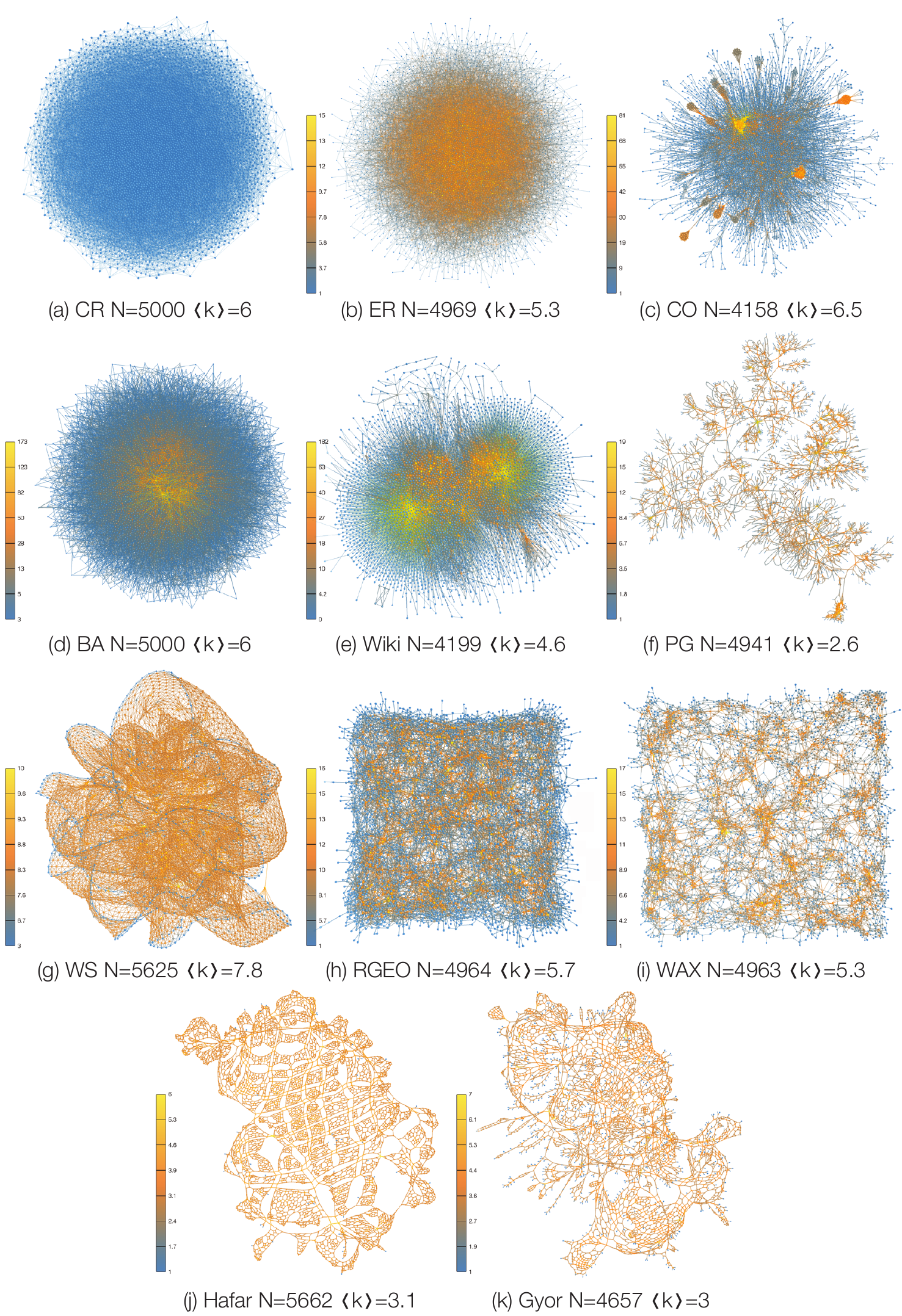} \\
  \caption{Visualizations of the adopted theoretical
  and real-world networks, with respective sizes
  and average degrees.  The position of the nodes, whenever available, was not taken into account in order to favor the visualization of the interconnecting topology.}
  ~\label{f:networks}
  \end{center}
\end{figure*}

The performed experiments are aimed at illustrating the estimation of the malleability with respect to the average shortest path length, clustering coefficient, degree assortativity, and degree entropy for a set of $11$ complex network topologies comprising both theoretic and real-world data described as following. Also, visualizations obtained for the considered networks are shown in Figure~\ref{f:networks}.

\begin{itemize}
\item {\bf Crystal model (CR)}: This kind of network is constructed by starting with $N$ disconnected nodes which are progressively connected to other randomly selected nodes, up to the point when exactly $k$ connections are obtained. The expected resulting network is perfectly regular with each node having degree $k$. Note that, for this technique to work, several trials may be needed to ensure that every node reaches $k$ connections.

\item {\bf Erd\H{o}s-Rényi model (ER)}: This corresponds to the traditional random network model~\cite{Albert:2002p161,Costa2011Survey}, in which each pair of nodes has the same probability of being connected.

\item {\bf Collaboration Network (CO)}: Scientific collaboration network for authors of papers submitted to arXiv covering the areas of General Relativity and Quantum Cosmology~\cite{leskovec2007graph}.

\item {\bf Barab\'asi-Albert model (BA)}: Traditional model of networks having scale-free behavior introduced by Barab\'asi-Albert~\cite{Albert:2002p161}. In this model, a network is grown in such a way that the probability of a node receiving a connection is proportional to its current degree.

\item {\bf Subset of the Wikipedia network (Wiki)}: This subset of the Wikipedia encompasses all articles categorized as being related to Brazil or Portugal. Each node corresponds to an article and a connection indicates a reference between them. For this analysis, the direction of the connections was disregarded. 

\item {\bf US Power Grid (PG)}: This network corresponds to the power grid infrastructure of the western states on the United States of America~\cite{watts1998collective}. Nodes represent distribution stations, transformers and generators, while edges correspond to the transmission lines.

\item {\bf Watts-Strogatz model (WS)}: From an initial regular network, each edge has a probability $p$ of being rewired. Even for small $p$, i.e., a few rewirings, the network starts to display small-world phenomena, in which the average distance between any nodes is dramatically reduced in comparison with the original regular network~\cite{Watts:1998kx}. Here we opted to start with a 2D regular network and $p=0.005$.

\item {\bf Random Geographic Network (RGEO)}: Corresponds to the traditional geographic model~\cite{Barthelemy2011Sp} in which nodes are distributed over a space and connected if the distance between then is smaller than a given threshold.

\item {\bf Waxman model (WAX)}: This is also a geographic model of complex networks~\cite{Barthelemy2011Sp}. However, the connections are established stochastically with a probability that depends on the distance between pairs of nodes.

\item {\bf Hafar urban network (Hafar)}: Urban network obtained for the city of Hafar in Saudi Arabia. Each node corresponds to a crossing or termination, and connections represent the streets. In particular, this network was obtained by using the recently introduced methodology to extract topologies of urban centers~\cite{comin2016diffusion}, being part of the dataset obtained in ~\cite{domingues2018topological}.

\item {\bf Gyor urban network (Gyor)}: Similar to the previous network but for the city of Gyor in Hungary.

\end{itemize}

Each experiment starts with a specific complex network, e.g.~a realization of the ER model containing $E$ edges.  
Then, edge $i=1$ is removed from the network, and the four considered topological properties are estimated for that new network.  Edge $i$ is then returned to the original network, and the following edge $i =2$ is removed, new measurements are obtained, and so on.

\section{Results}
\label{s:results}

Table~\ref{table:malleabilityTable} shows the four topological measurements obtained for the original networks, as well as the respective malleabilities.  The last column shows the maximum malleability for each network.

\begin{table*}[!htbp]
\begin{tabular}{r|cccc|ccccc}
\multicolumn{1}{l|}{} & \multicolumn{4}{c|}{Original measurement} & \multicolumn{5}{c}{Malleability} \\
\multicolumn{1}{l|}{Network} & \multicolumn{1}{c}{$\langle L \rangle$} & \multicolumn{1}{c}{$C_c$} & \multicolumn{1}{c}{$A_d$} & \multicolumn{1}{c|}{$E_d$} & \multicolumn{1}{c}{$M(\langle L \rangle)$} & \multicolumn{1}{c}{$M(C_c)$} & \multicolumn{1}{c}{$M(A_d)$} & \multicolumn{1}{l}{$M(E_d)$} & \multicolumn{1}{l}{Estimated} \\ \hline
WS & 12.4 & 0.432 & 0.0019 & 0.54 & 102 & 11 & 14 & 2.3 & 102 \\
CR & 5.2 & 0.001 & 0.0000 & 0.00 & 794 & 1.1 & 1.0 & 1.0 & 794 \\
RGEO & 31.8 & 0.581 & 0.0663 & 2.22 & 101 & 1076 & 1234 & 42 & 1234 \\
PG & 19.0 & 0.080 & 0.0008 & 1.70 & 2614 & 17 & 577 & 35 & 2614 \\
ER & 5.3 & 0.001 & 0.0003 & 2.22 & 2988 & 1.3 & 1250 & 48 & 2988 \\
CO & 6.0 & 0.457 & 0.2034 & 2.66 & 389 & 2785 & 3412 & 317 & 3412 \\
WAX & 12.7 & 0.093 & 0.0136 & 2.23 & 3627 & 270 & 1627 & 49 & 3627 \\
Gyor & 28.5 & 0.077 & 0.0043 & 0.99 & 4309 & 5.1 & 30 & 4.4 & 4309 \\
Hafar & 29.1 & 0.012 & 0.0033 & 0.47 & 4989 & 1.3 & 7.3 & 2.3 & 4989 \\
BA & 4.1 & 0.007 & -0.0231 & 2.05 & 2616 & 20 & 9681 & 116 & 9681 \\
Wiki & 2.8 & 0.522 & -0.0653 & 3.01 & 84 & 10456 & 12253 & 355 & 12253 \\ \hline
\end{tabular}
 \caption{\label{table:malleabilityTable} Measurements and respective malleability for the considered networks. The last column indicates the maximum malleability obtained for each network.}
\end{table*}

A first interesting result concerns the confirmation of the fact that the malleability for a given network can depend significantly on the chosen measurement.  For instance, the Wiki network has the smallest $M_L$ and largest $M_{A_d}$.  Regarding the maximum malleability, it was observed for the shortest path in 7 of the considered networks, while the other 4 cases implied largest malleability for $A_d$.  The network with the smallest malleability corresponds to the $WS$.  This network is almost identical to a lattice, except for a few rewirings.  As such, the removal of distinct edges will lead to isomorphic new network configurations, hence the small malleability.  The crystal network is embedded in a much higher dimensional space than the aforementioned WS, having a border (corresponding to the latest edge additions along the network construction), and therefore a center.  As such, edge removals at different positions of this network can produce new non-isomorphic network configurations, hence the second smallest malleability observed for this structure.  

The largest malleability was obtained for the Wiki network, followed by BA structures and the two considered cities (Gyor and Hafar).  Interestingly, the maximum malleability in the two latter networks (cities) correspond to the average shortest path length, while the degree assortativity led to the highest malleability in the case of the BA and Wiki networks.  In the case of the two cities, which are spatial structures, the average shortest path length can change considerably as an edge is removed, because there are relatively few shortest paths with the same length interconnecting a pair of nodes.  Therefore, edge perturbations tend to yield a relatively large number of new configurations of the original network, implying in higher malleability.  Contrariwise, in the case of the BA and Wiki networks, which are small world structures, there is a relatively larger number of shortest paths with the same length interconnecting two points, ultimately leading to smaller malleability for this measurement.

\section{Concluding Remarks}
\label{s:conclusions}

We have suggested a measurement, namely the \emph{malleability}, to quantify the potential of obtaining new topological configurations as a consequence of incremental modifications of a complex network, such as those observed with the growth of the network or from attacks.  In order to avoid the identification of isomorphisms, we resource to considering the malleability of networks with respect to specific topological measurements such as shortest path length and assortativity.  Though the mapping of network configurations into a single measurement is, generally, not one-to-one, this measurement-based approach to malleability allows the discussion of the potential of a network to produce new configurations from the point of view of specific properties such as shortest path length and assortativity.  

The obtained results indicate that the malleability depends on specific measurements, with the shortest path length and degree assortativity producing the largest malleability values.  We also found that the Wikipedia network yielded the largest malleability (with respect to degree assortativity), while the smallest malleability was obtained for the Watts-Strogatz model. 

\section*{Acknowledgements}
Filipi N. Silva thanks FAPESP (grant no. 2015/08003-4 and 2017/09280-7) for sponsorship. Cesar H. Comin thanks FAPESP (grant no. 15/18942-8) for financial support. Luciano da F. Costa thanks CNPq (grant no. 307333/2013-2) and NAP-PRP-USP for sponsorship. This work has been supported also by FAPESP grants 11/50761-2 and 2015/22308-2 and CAPES. Research carried out using the computational resources of the Center for Mathematical Sciences Applied to Industry (CeMEAI) funded by FAPESP (grant 2013/07375-0).

\bibliographystyle{unsrt}
\bibliography{references}

\appendix
\section{Precision analysis for the malleabilities}

\begin{figure*}[!htbp]
 \centering
\subfigure[]{\includegraphics[width=0.45 \linewidth]{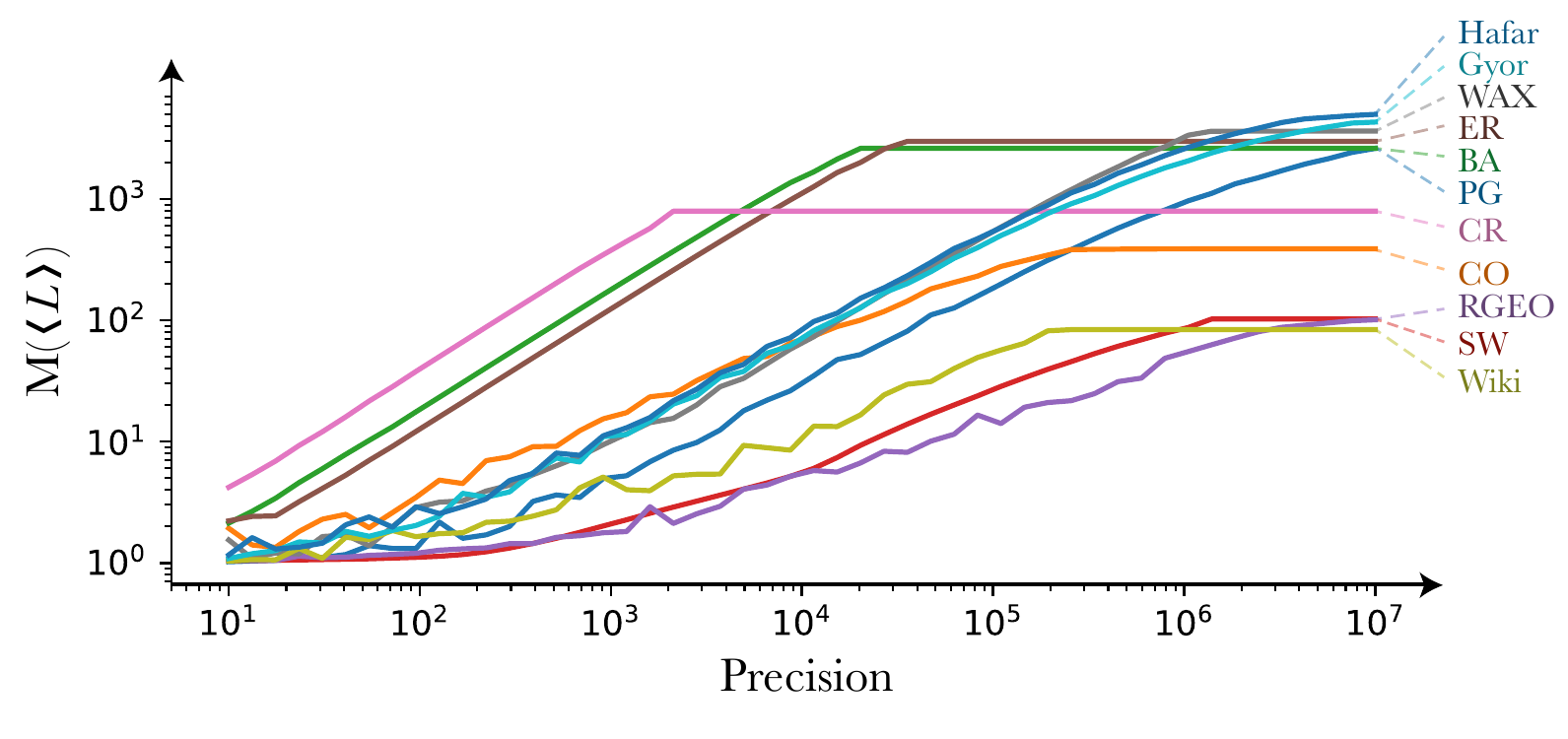}}~
\subfigure[]{\includegraphics[width=0.45 \linewidth]{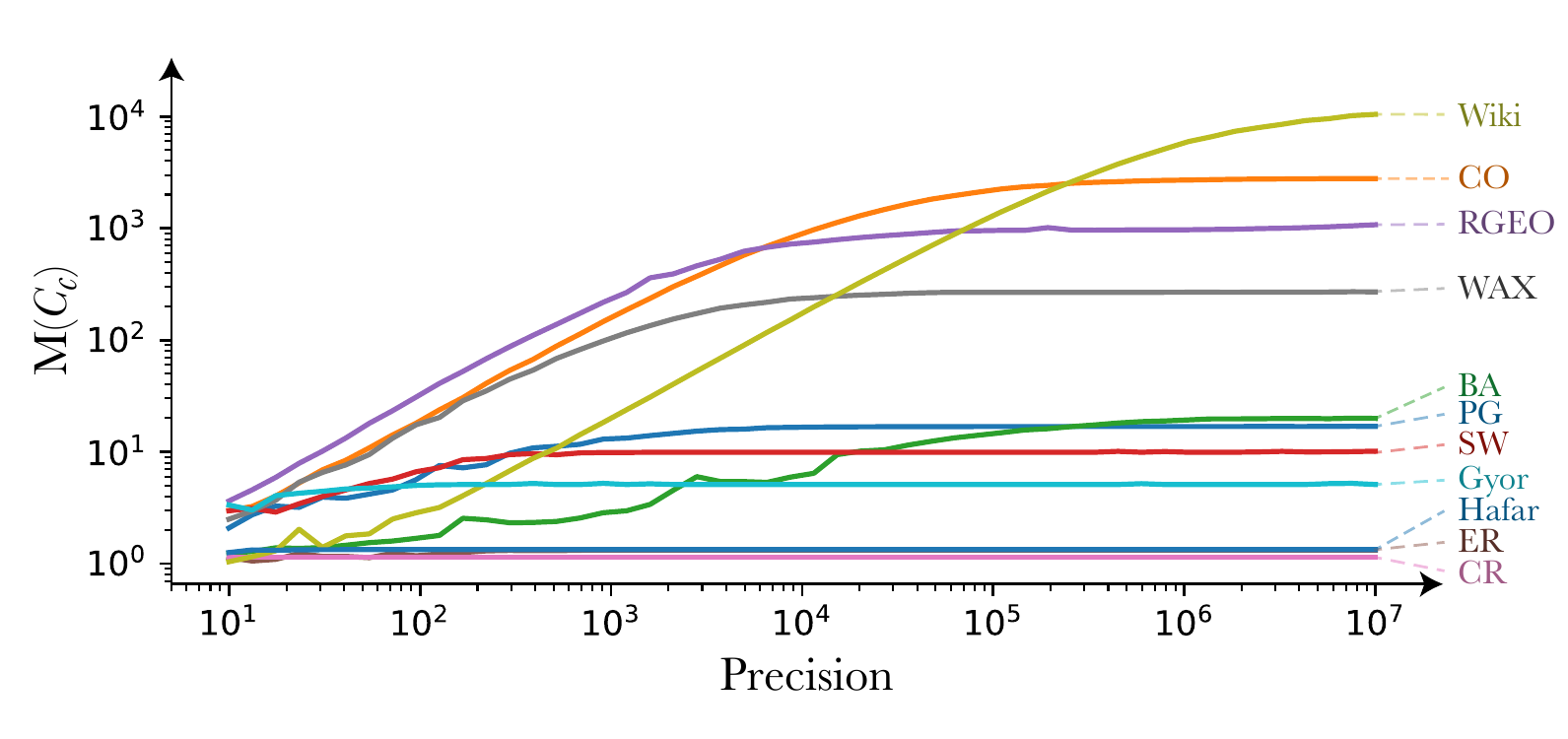}}\\
\subfigure[]{\includegraphics[width=0.45 \linewidth]{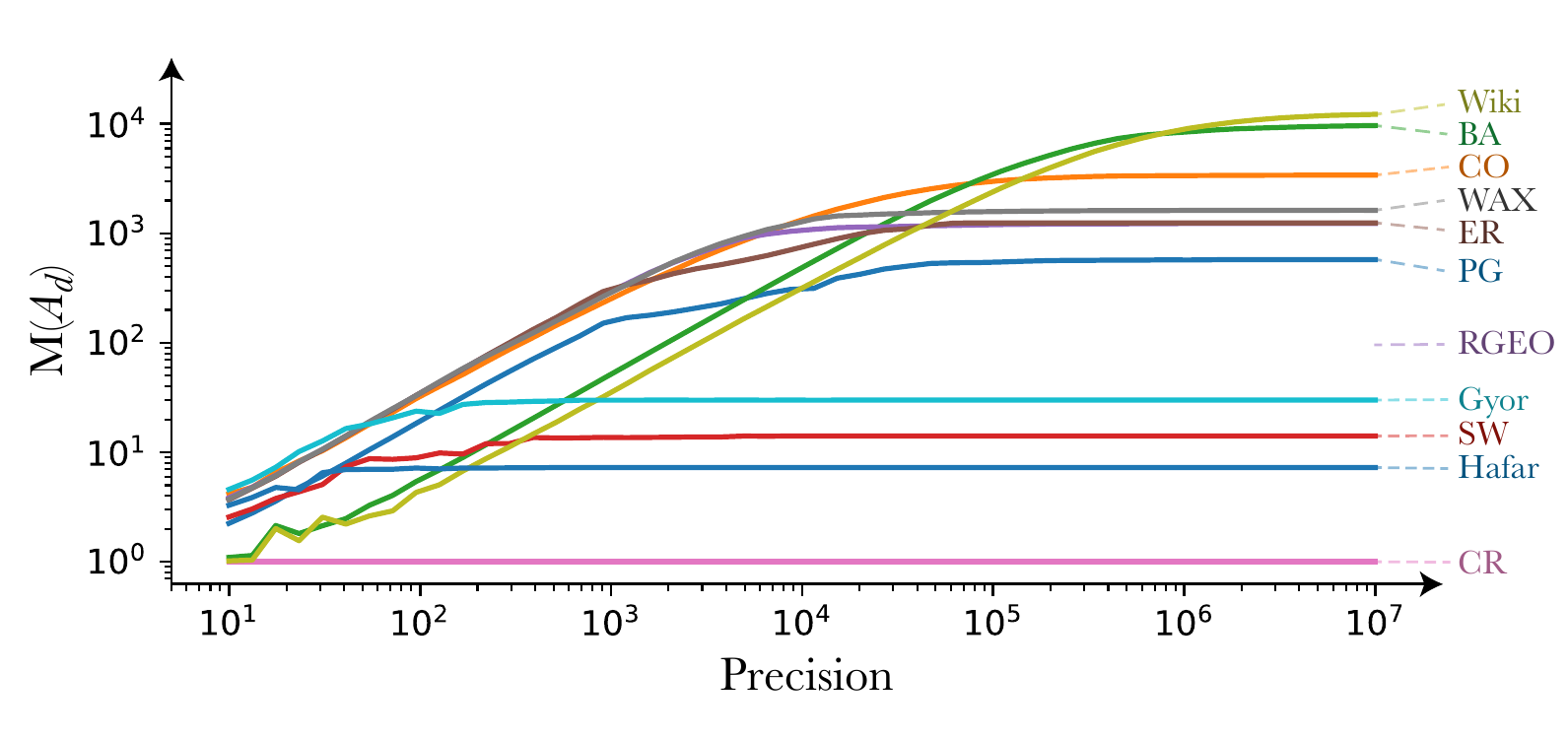}}~
\subfigure[]{\includegraphics[width=0.45 \linewidth]{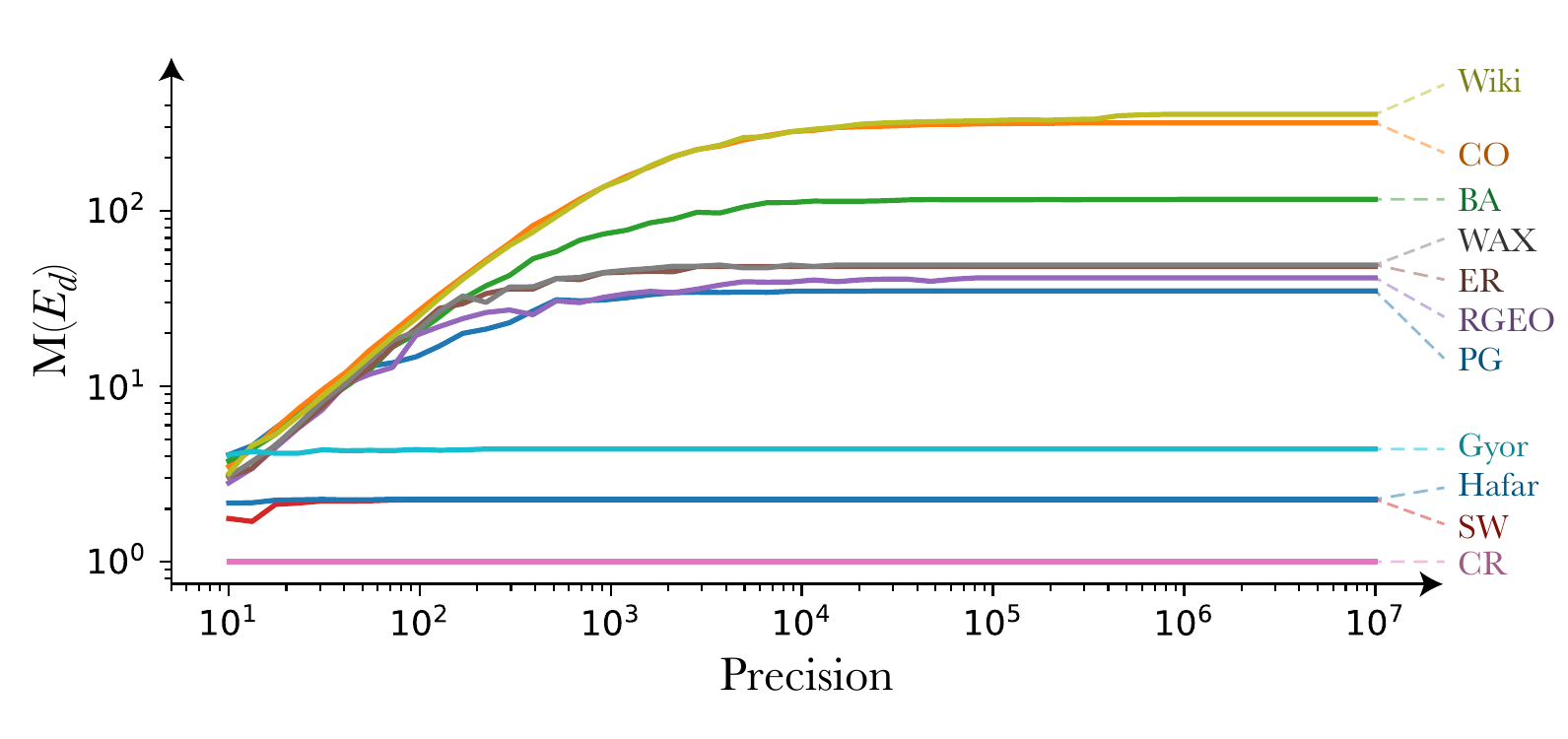}}
\caption{\label{fig:maleabilityPrecision} Malleability vs precision.}
\end{figure*}

Because of the discrete nature of the computational representation of measurements, we also investigated the effects of the precision of their computational representation on the respective malleabilities. In order to do so, we obtained the malleability considering increasing levels of representation precision. The results are shown in Figure~\ref{fig:maleabilityPrecision}. We found that the malleabilities obtained for the majority of the measurements are stable for precisions above $10^6$, corresponding to about $6$ digits. The exception are the malleabilities obtained for the average shortest path length, which are stable only after $7$ digits.  

\end{document}